\documentclass[a4paper,10pt]{article}
\usepackage[utf8]{inputenc}
\usepackage[utf8]{inputenc}
\usepackage{amsmath}
\usepackage{graphicx}
\usepackage{subfig}
\usepackage{color}
\usepackage{amssymb,amsmath,amsfonts}
\usepackage{graphicx}
\usepackage{epstopdf}
\usepackage{bm}
\usepackage{latexsym}
\usepackage{graphics, setspace}
\usepackage{color}
\newcommand{\be}{\begin{equation}}
\newcommand{\ee}{\end{equation}}
\newcommand{\bea}{\begin{eqnarray}}
\newcommand{\eea}{\end{eqnarray}}
\newcommand{\<}{\langle}
\renewcommand{\>}{\rangle}

\def\<{\langle}
\def\>{\rangle}
\def\(({\left(}
\def\)){\right)}
\def\[[{\left[}
\def\]]{\right]}

\begin{document}

\title{Parallel Tempering for the planted clique problem}
\author{Angelini Maria Chiara\\
\scriptsize Dipartimento di Fisica, Ed. Marconi, "Sapienza" Universit\`a di Roma, P.le A. Moro 2, 00185 Roma Italy}
\date{}

\maketitle

\begin{abstract}
The theoretical information threshold for the planted clique problem is $2\log_2(N)$, but no polynomial algorithm is known to recover a planted clique of size 
$O(N^{1/2-\epsilon})$, $\epsilon>0$. In this paper we will apply a standard method for the analysis of disordered models, the Parallel-Tempering (PT) algorithm, 
to the clique problem, showing numerically that its time-scaling in the hard region is indeed polynomial for the analyzed sizes. 
We also apply PT to a different but connected model, the Sparse Planted Independent Set problem. In this situation thresholds should be sharper and 
finite size corrections should be less important. Also in this case PT shows a polynomial scaling in the hard region for the recovery.
\end{abstract}

\section{Introduction}
The planted clique problem is the following \cite{plantedclique}:
We extract a random graph with $N$ nodes, each node is connected to another one with probability $p$.  
Then, we plant  a clique $\mathcal{C}$, imposing $K$ nodes among the $N$ ones to be connected to each other with probability
$q=1$. Given the resulting graph, we search for an algorithm able to identify the elements of the planted clique. 
The problem can be studied for all the values of the probabilities $p$ and $q$, 
but in the following we will focus on the values $p=\frac{1}{2}$, $q=1$ \footnote{Clearly, if $q\neq1$, $\mathcal{C}$ is no more a real clique, but the problem 
to find it retains the same properties and can be analyzed with the same methods of the $q=1$ case.}.
Given a random graph with probability $p=\frac{1}{2}$ to have a link between two nodes, 
one can easily show that the largest purely random clique is of size $K_{max}(N)\simeq2\cdot\log_2(N)$ for
$N$ large \cite{GM75}.  
As a consequence, an exhaustive search for a clique of size $K$ returns the planted clique if $K>K_{max}$.
However, it has long been conjectured that no polynomial-time algorithm can find cliques of size $N^{1/2-\epsilon}$, $\epsilon>0$.
In Ref. \cite{Barak} this has been proved for the class of sum-of-squares algorithms. 
In Ref. \cite{Montanari}, a message passing algorithm has been constructed that fails unless $K>K_{BP}=\sqrt{\frac{N}{e}}$.
In Ref. \cite{Montanari15}, an analogous gap between the threshold for exhaustive and polynomial algorithms has been found in the sparse clique problem.
In this case, it is shown how the problem undergoes two phase transitions: The first one is a dynamical one, and below that threshold no local algorithm
is able to find the planted clique. The second transition is a static transition that identifies the threshold for exhaustive search.
The existence of such kind of gaps
is common in recovery problems, but the clique problem is
in a certain manner special because for this problem the two thresholds scale differently with $N$. 
Some recent works prove hardness in other problems assuming the hardness
of the planted clique problem (in the corresponding region) \cite{hard1},\cite{hard2},\cite{hard3},\cite{hard4}. For this reason, there is a lot of current interest in the planted clique problem.
In the first part of this paper we will show that also in this dense limit, the clique problem undergoes a static and a dynamic phase transition, 
as demonstrated in \cite{Montanari15} in the sparse case.
Then applying standard methods for the analysis of disordered models, and in particular a Parallel Tempering (PT) algorithm, we will see how to find the planted clique down to $K_{max}$.
The thermal algorithms are not easy to be analyzed: In particular, the challenge is to understand how the time of convergence scales with $N$.
We show that data from PT are in very good agreement with a polynomial scaling.
However, for the clique problem an exhaustive search algorithm can find the planted clique of size $K>2\log_2(N)$ in a time $O(\exp(c\log^2(N)))$:
It is sufficient to find a clique of size $k_0 = 2\log_2(N)$, that takes a time $\binom{N}{k_0}$ and then to expand starting from that one.
Thus it is quite difficult to distinguish between a polynomial or a non-polynomial $O(\exp(c\log^2(N)))$ behavior. 
For this reason, in the second part of this paper, we move to a different but connected model: The planted Independent Set (IS) model. 
Being this problem sparse, the thresholds are sharper:
The exhaustive algorithm can find solutions in time $O(\exp(N))$. For this problem in the hard region we numerically show that for the analyzed sizes the PT algorithm can find solutions
in polynomial time. This gives good reasons to believe that also in the case of the planted clique problem the PT algorithm finds solutions in polynomial time.
\section{The planted clique problem and its transitions}
To be concrete, we construct a graph of $N$ nodes with a planted clique $\mathcal{C}$ of size $K$. 
On each node there is a variable $v_i$, $v_i=1$ if node $i\in\mathcal{C}$, $v_i=0$ if node $i\notin\mathcal{C}$. 
On each edge between nodes $i$ and $j$ we put a variable  $A_{ij}$. The edge variables are 0 or 1 with the following probabilities:
\begin{equation}
p(A_{ij}=1|    \{{v\}})=
\begin{cases}
 1  &\text{if } v_iv_j=1 \\
 \frac{1}{2} &\text{otherwise}
\end{cases}.
\label{eq:likelihood}
\end{equation}
Given a realization of the graph we want to estimate the variables $v_i$. We will call our estimation $x_i$.
We introduce a Belief-Propagation (BP) algorithm that is essentially the one proposed in Ref. \cite{Montanari}.
Following the Bayes formula, the posterior probability for $x_i$ given the graph is
\begin{equation}
P(x_i|\{{A_{ij}\}})=P(A_{ij}|\{{x_i\}})P(x_i),
\label{eq:posterior}
\end{equation}
where the likelihood $P(A_{ij}|\{{x_i\}})$ is the one of eq. (\ref{eq:likelihood}) and $P(x)$ is the \textit{prior} probability. 
The original problem has a global constraint on the size of the clique to be recovered (we are treating the case of known $K$):
$P(\{{x\}})$ should be zero if $\sum_i x_i\neq K$.
However we are using local algorithms, that cannot implement global constraints. 
Thus we choose a local prior on the single node: $P(x)=\left(\frac{K}{N}\right)^{x}\left(1-\frac{K}{N}\right)^{1-x}$.
The BP algorithm is a way to extract the marginal probabilities for each node from Eq. (\ref{eq:posterior}).
We introduce cavity messages $\psi_{i\rightarrow j}(x_i)$ proportional to the probability that node $i$ takes value $x_i$,
conditioned on the absence of edge $(ij)$. 
Iterative equations on the messages read:

\begin{align*}
 \psi_{i\rightarrow j}(x_i=0)=&\frac{N-K}{N}\((\frac{1}{2}\))^{N-1},\\
 \psi_{i\rightarrow j}(x_i=1)=&\frac{K}{N}\((\frac{1}{2}\))^{N-1}\prod_{k\backslash j}\[[1+(2A_{ij}-1)\psi_{k\rightarrow i}(1) \]].
\end{align*}

Cavity probabilities are obtained from the normalization of the cavity messages:
$\eta_{i\rightarrow j}(x_i)=\frac{\psi_{i\rightarrow j}(x_i)}{z_{i\rightarrow j}}$, with $z_{i\rightarrow j}=\psi_{i\rightarrow j}(0)+\psi_{i\rightarrow j}(1)$.

Once the iteration of the cavity messages has reached a fixed point, marginal probabilities are obtained as:
\begin{align}
 \psi_{i}(x_i=0)=&\frac{N-K}{N}\((\frac{1}{2}\))^{N},\\
 \psi_{i}(x_i=1)=&\frac{K}{N}\((\frac{1}{2}\))^{N}\prod_{k}\[[1+(2A_{ij}-1)\psi_{k\rightarrow i}(1) \]].
\end{align}
$\eta_{i}(x_i)=\frac{\psi_{i}(x_i)}{z_{i}}$, with $z_{i}=\psi_{i}(0)+\psi_{i}(1)$.
The \textit{Bethe free energy} associated to the reached solution is $f=-\frac{1}{N}\[[\sum_i\log(z_i)-\sum_{ij}\log(z_{ij})\]],$
with $z_{ij}=\frac{z_i}{z_{i\rightarrow j}}$.
We then assume the elements of the clique to be the first $K$ elements with largest $\eta_i(1)$.
In the thermodynamic limit the recovery is possible if $K>K_{BP}=\sqrt{N/e}$ \cite{Montanari}.
Let us emphasize the difference in the Fixed Point of BP when the algorithm finds the planted clique and when it does not:
If the planted clique is recovered, the BP messages are completely polarized: $\eta_i(1)=0$ (if $i\notin\mathcal{C}$) or $\eta_i(1)=1$ (if $i\in\mathcal{C}$).
If the planted clique is not recovered, the fixed point $\eta^*_i$ reached by BP is not completely polarized: 
$\eta_i(1)=\eta^*_i(1)\neq0,1$ $\forall i$. We will call this solution the \textit{paramagnetic solution}.

In the rest of this Section, we will identify the different phases of the planted clique problem.
We will introduce the scaling variable $\tilde{K}=K/\log_2(N)$, to take into account the fact that the important thresholds of the model scale non-trivially with the size of the system.
In terms of this variable, we will see that the finite-size scaling reduces to the usual one in presence of a first-order phase transition.

\begin{figure}
\centering
\includegraphics[width=0.7\textwidth]{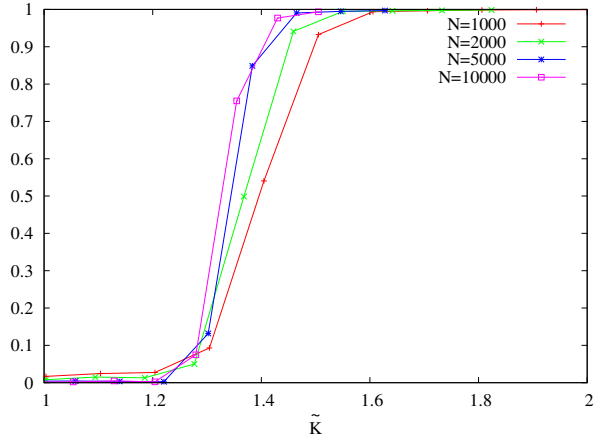}
\caption{\label{Fig:planted_stability}Probability for recovery starting around the planted solution. Averages are performed over 100 samples. Even if we know that random cliques of size 
$\tilde{K}\leq \tilde{K}_{ran}=2$ are present, the planted solution
remains locally stable down to $\tilde{K}_{sp}\simeq1.3$. }
\end{figure}

First of all, we study the stability of the planted solution. To do this, we initialize the BP messages near enough to the planted solution
and we look to the solution reached after iteration. Results are shown in Fig. \ref{Fig:planted_stability}.
The planted solution is reached well below the threshold $\tilde{K}_{ran}=2$, until the threshold $\tilde{K}_{sp}\simeq1.3$.
This value is extracted from the intersection of data in Fig. \ref{Fig:planted_stability} for the two larger analyzed sizes, $N=5000$ and $N=10000$.
We know that there exist other random cliques of size $K\le K_{max}$, but for $K_{sp}<K<K_{max}$ the planted solution is still locally stable under small perturbations:
Speaking in terms of the free energy, the planted solution represents a local minimum, well separated from the paramagnetic one. 
In the statistical physics language, $K_{sp}$ corresponds to the \textit{spinodal} point for the existence of the planted solution.
\footnote{Below the spinodal of the planted solution, in general, the recovery in the non-planted ensemble is easy \cite{QuietPlanting}. 
This should imply that for $K<K_{sp}$
there should exist a polynomial algorithm able to find a random clique of size $K$. 
However, the threshold for the polynomial algorithms in the non-planted case is believed to be $\tilde{K}_{Karp}=1$ 
in the large $N$ limit \cite{Karp}. Whether the difference between $\tilde{K}_{sp}$ and $\tilde{K}_{Karp}$ 
is just due to finite size effects or has a deeper meaning should be better analyzed and will
be the subject of a subsequent work.}

Then, we compare the average Bethe free energy $\overline{f}_{plan}$ and $\overline{f}_{ran}$ of the solutions found respectively from planted and random initialization, 
where the average is intended over different realizations of the planted clique and of the graph at fixed $N$ and $K$.
We name as $\tilde{K}_s(N)$ the threshold at which $\overline{f}_{ran}(N)=\overline{f}_{plan}(N)$; it corresponds to a \textit{static} phase transition, at which
the global minimum of the free energy changes from the planted to the paramagnetic solution. 
\begin{figure}
\centering
\includegraphics[width=0.7\textwidth]{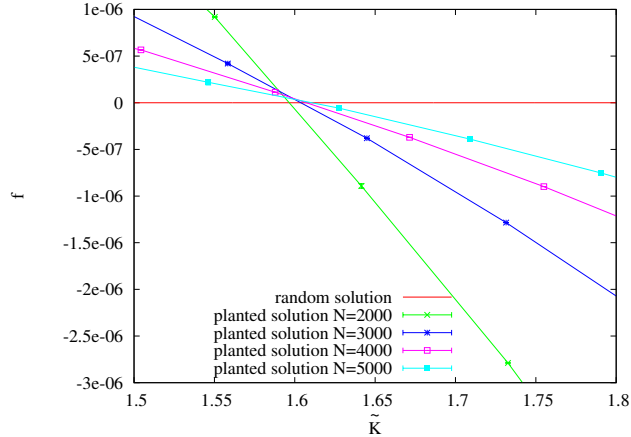}
\caption{\label{Fig:f}Comparison between the average free energy $f$ of the paramagnetic solution and the one of the planted solution. Averages are over 10 samples.
The point where the planted solution has the same $f$ with respect to the paramagnetic solution locates the static threshold $\tilde{K}_s$.}
\end{figure}
For large $N$, it was analytically shown that $\tilde{K}_s(\infty)=2$ \cite{Castro_Ks}. We find, however, that finite size corrections are huge and for finite $N$, $\tilde{K}_s(N)$ 
is quite different from its large $N$ limit: $\tilde{K}_s(10000)\simeq1.6$ as shown in Fig. \ref{Fig:f}.
The values found for $\tilde{K}_s(N)$ are in good agreement with the finite size estimate for the largest size $K_{max}(N)$ of a random clique in a random graph of size $N$.
$K_{max}(N)$ can be obtained by the following probabilistic argument:
The expected number of cliques of size $K$ in an Erd\"os-R\'enyi graph of size $N$ with bonds present with probability $p$ is given by
$E(N,K) = \binom{N}{K} p^{\binom{K}{2}}.$
Defining $K_{max}(N)$ as the largest integer $K$ for which $E(N,K) > 1$, 
the largest naturally occurring clique is shown to have with high probability size $K_{max}$ or $K_{max}+1$ in graphs with $N$ large \cite{Matula1,Matula2,Bollobas}.
$K_{max}(N)$ asymptotically tends to $2\log_2(N)$ for $N\rightarrow\infty$. Thus in the case of the planted clique we find that the static threshold
corresponds to the threshold for the existence of purely random cliques.

Summarizing:
\begin{itemize}
 \item For $K>K_{BP}$ there is just one minimum of the free energy, that is the one corresponding to the planted clique, found by randomly initialized BP. 
We will call this phase an \textit{easy} phase for the recovery.
\item For $K_{s}<K<K_{BP}$ there are two local minima of $f$: The global one corresponds to the planted clique,
and it is reached by BP with planted initialization; the other
local minimum corresponds to a paramagnetic solution, that is the one where BP stops if randomly initialized. 
If we know how to nucleate the planted solution, it is still possible to find it. We will call this phase a \textit{hard} phase.
\item For $K_{sp}<K<K_s$ the planted solution is still a local minimum of the free energy, that can be found if BP is initialized around the planted solution. 
However, the global minimum of $f$ is the paramagnetic one. This phase corresponds to an \textit{impossible} phase.
\item For $K<K_{sp}$ the number of random cliques is large and the planted solution is no more stable, it is no more a minimum of the free energy: 
Even starting near to the planted solution, the BP algorithm will flow to the unique paramagnetic minimum.
\end{itemize}
A similar analysis was performed in Ref. \cite{clusteringDecelle} for 
the so-called stochastic block model
and in Ref. \cite{Montanari15} for the problem of finding a highly connected subset of vertices
in a sparse graph. From the results of Ref. \cite{Montanari15}, the dynamical threshold associated with the failure of BP in the sparse case is shown to reduce to 
$K=\sqrt{N/e}$ in the dense planted clique case, while the thresholds associated with the static and the spinodal phase transition 
are located at $K^*=o(\sqrt{N})$ in the dense limit.
In statistical physics such a situation is called a first order transition: There are two competing minima of $f$; varying the parameters of the problem, 
the global one changes from one to another.
Such a situation is present in other recovery problems. A well-known example is the compressed sensing \cite{CS}.

\section{Parallel Tempering for the planted clique problem}
\begin{figure}
\centering
\includegraphics[width=0.48\columnwidth]{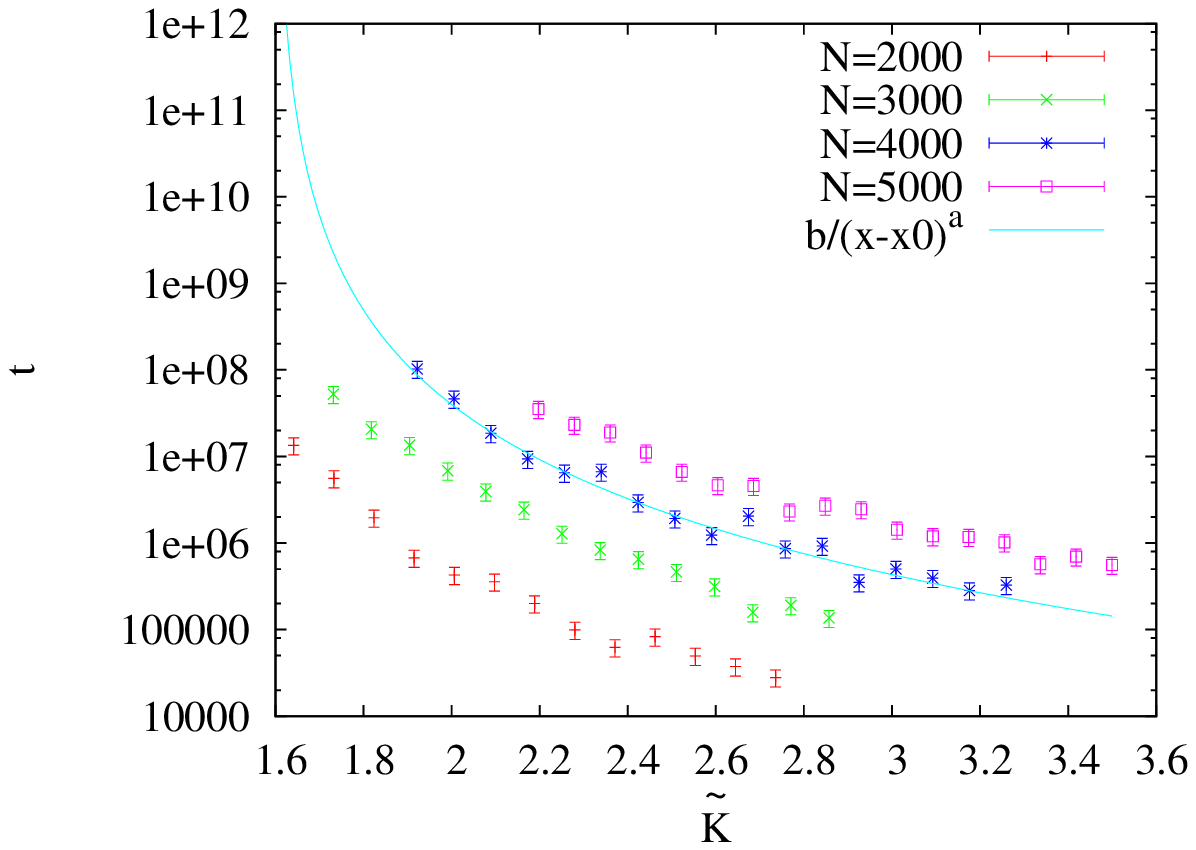}
\includegraphics[width=0.48\columnwidth]{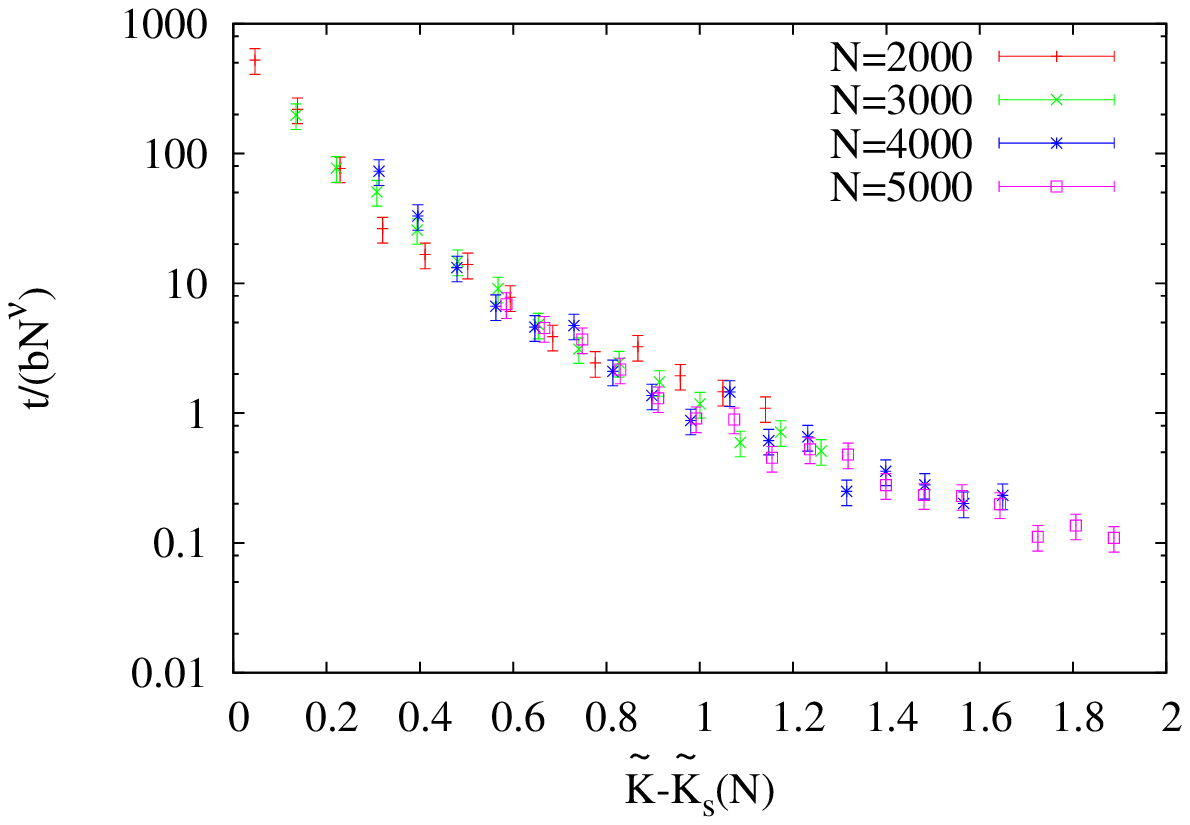}
\caption{\label{Fig:tempiPT}Left: Time of convergence for PT changing the size $K$ of the planted clique, and the size of the graph.
Each point is the average over 20 realizations of the graph. 
The time of convergence diverges as a power-law at $\tilde{K}_s(N)$. 
Right: Collapse of data for the convergence time once proper rescaled variables are used. 
The time of convergence grows as a power law with $N$.}
\end{figure}

We will now look to the statistical mechanical model constructed by introducing a Hamiltonian associated to the Bayes probability for the clique problem,
following an approach similar to Ref. \cite{Hu}.
In statistical physics, the appearance of metastable minima is encountered in a large number of problems. 
In these cases, the more efficient algorithm for the research of the true minimum is the so-called \textit{Parallel Tempering} (PT) \cite{PT}, that we will apply to the clique problem.
We define an Hamiltonian associated to the posterior probability of eq. (\ref{eq:posterior}) as: $P_{\beta}(\{x\}|\{{A\}})\equiv e^{-\beta H}$,
where the Hamiltonian has the form:
\small
\begin{align*}
H(\{{x\}})=&-\sum_i \log(P(x_i)) +\\
&-\sum_{ij}\[[\((1-A_{ij}\))\log\frac{\((1-x_ix_j\))}{2}+A_{ij}\log\frac{\((1+x_ix_j\))}{2}\]]
\end{align*}
\normalsize
The energy assumes the value $H=\infty$ if $A_{ij}=0$ and $x_ix_j=1$, preventing from having configurations without links between two elements of a clique.
We have introduced an additional parameter, the inverse temperature $\beta=\frac{1}{T}$, that takes the value $\beta=1$ in the original problem. 
Given a realization of the graph $\{{A\}}$, we introduce $n=19$ replicas of the system with the same graph, each replica $i$ is at a different inverse temperature: $\beta_i=1-i\cdot0.05$, $i\in[0,18]$. 
The $i=0$ replica is the original system.
For each replica we perform a standard Metropolis Monte Carlo (MC) simulation. After 5 MC steps for each replica, we try to flip the configuration of the $i$-th and $(i+1)$-th
replicas with probability 
\begin{equation}
p=\text{min}\((1,e^{\((\beta_i-\beta_{i+1}\))\((E_i-E_{i+1}\))}\)),
\label{eq:flipping}
\end{equation}
where $E_i$ is the actual energy value of the $i$-th replica. 
We then measure the magnetization $M$ of the original replica $i=0$: $M\equiv\sum_j x_j$. When $M=K$ we stop our simulation, having identified the 
planted clique.
If $K<K_{BP}$, the original system will be firstly attracted to the secondary minimum of the free energy and should overcome a barrier to reach the true minimum.
However at $\beta<1$, this barrier lowers or eventually it disappears. Replicas at higher temperature are free to explore a larger part of the phase space in less time.
Flipping replicas thus permits the original system to reach the true minimum in a smaller time.
Being the system fully connected, one could naively think that the MC algorithm takes $O(N^2)$ time:
An iteration step is intended as the attempt to flip each of the $N$ variables, and the computation of the new energy is of order $O(N)$.
However, the proposal to flip a spin is accepted only $O(K)$ times, because if $x_i=0$ and we propose to flip it, the flip is accepted only if $A_{ij}=1$ for all spins $j$ with $x_j=1$,
while if $x_i=1$ it is always possible to flip it. A single step of the algorithm thus is $O(K\cdot N)$. 
In Ref. \cite{Montanari15} it is stated that no local algorithm can find the planted solution for $K<K_{BP}$. PT is not local
because of the flipping procedure between replicas.
It finds the planted solution for $\tilde{K}>\tilde{K}_s(N)$.
Please note that the PT algorithm finds the planted solution until the actual size-dependent static threshold $\tilde{K}_s(N)$, that is well below 
the $N\to\infty$ static threshold $2\log_2(N)$ for the analyzed sizes. 
As in all first-order phase transitions, the time for convergence is diverging at the static transition point $\tilde{K}_s$.
The time of convergence seems to diverge as $t(N,\tilde{K})\propto\frac{N^{\nu}}{\left(\tilde{K}-\tilde{K}_s(N)\right)^a}$, as shown in Fig. \ref{Fig:tempiPT}.
The value for the exponents are $\nu=5.78(4)$, $a=3.64(12)$.
In the right part of fig. \ref{Fig:tempiPT} we show the collapse of the data for the convergence time as a function of $\tilde{K}$ 
for different sizes $N$ once the proper scaling variables are used.
Summarizing, the time of convergence grows as a power law with $N$ and diverges with a power law at the static threshold $\tilde{K}_s(N)$.
The value for the critical exponents $a$ and $\nu$ is not optimal. In fact, the parameters of PT (number of replicas, spacing in temperature...) can be modified 
leading probably to a changing in the exponents. However, our purpose is not to optimize the PT but only to show that it is indeed a polynomial algorithm.
For the clique problem, an exhaustive search algorithm can find the planted clique in a time $O(\exp(c\log^2(N)))$. 
Even if we showed that the collapse of the data with a polynomial scaling is very good, 
one could always criticize that the analyzed sizes are too small to capture the difference between a polynomial and an exponential $O(\exp(c\log^2(N)))$ behavior.
Thus a better setting to numerically study the problem is a sparse graph, for which thresholds are sharper and a non-polynomial algorithm
should take $O(\exp(cN))$ time. In this situation, it is easier to distinguish between a polynomial or an exponential algorithm.
A first way to translate the planted clique problem on a sparse graph is the one proposed in Ref. \cite{Montanari15}.
A second way is to move to the planted Independent Set problem, as explained in the next Section.
We will show that also for this problem in the hard region the scaling of the PT time to find solutions
is well-fitted by a power-law for the analyzed sizes. This gives indications to believe that also in the case of the planted clique problem 
the PT algorithm really finds solutions in polynomial time.

\section{The planted independent set problem}
An Independent Set (IS) is a subset of vertices of a graph that are not connected. 
It is clear that a clique becomes an Independent Set on the complementary graph, thus the original problem to find a planted Clique results in finding an IS on the complementary graph.
The useful property of the IS problem is that it can be safely defined on a sparse graph of finite degree $d$
\footnote{The original dense problem to find the planted clique as defined in the previous section is exactly equivalent to the problem of finding a planted IS when $d=N/2$}.

We define the density of the IS $\rho=\frac{K}{N}$.
For $d>30$, one can show that the paramagnetic solution of the BP equations to find a random IS of size $\rho$ is stable up to density $\rho_l(d)$ well above the 
density $\rho_{max}(d)$ of the largest random IS \cite{BarbierIS}. This means that if we study the planted problem, we expect the existence of a region
in the density of the planted IS $\rho_{max}(d)<\rho(d)<\rho_l(d)$ where it will be hard
to find the planted IS even if we are above the theoretical threshold for the recovery: We are in a situation analogous to the case of 
the planted clique of size $K_s<K<K_{BP}$.
The thresholds for the IS problem as a function of $d$ on a Random Regular Graph has been computed in Ref. \cite{BarbierIS}.

\begin{figure}
\centering
\includegraphics[width=0.7\textwidth]{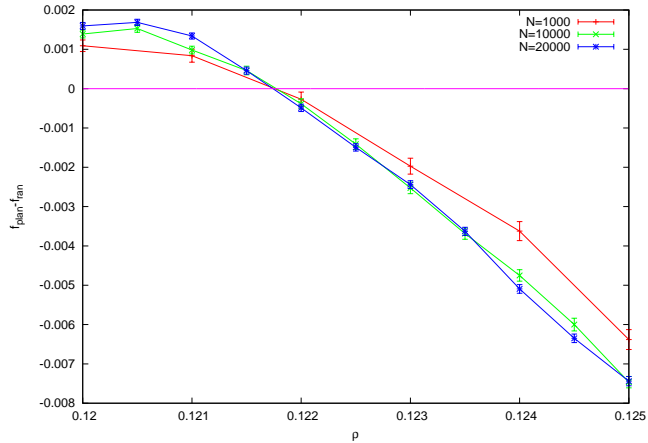}
\caption{\label{Fig:f_IS}Comparison between the free energy $f$ of the paramagnetic solution (reached by BP randomly initialized) and the one of the planted solution 
(reached by BP initialized near enough to the planted solution) for the planted IS problem on a Bethe lattice with average degree $d=40$.
The point where the planted solution has the same $f$ with respect to the paramagnetic solution locates the static threshold $\rho_s(40)=0.1217(1)$. Averages are performed over 1000 sample for $N=1000$
and 300 samples for $N=10000,20000$.}
\end{figure}

To be concrete, we plant a IS ${\cal I}$ of size $K$ on a graph of size $N$ with average degree $d$ in the following way: 
We extract link $A_{ij}=1$ between nodes $i,j\notin {\cal I}$ with probability $c_{in}$, and links between  $i\in {\cal I},j\notin {\cal I}$ with probability $c_{out}$.
We do not put links between $i,j\in\cal{I}$.
Imposing that all elements have average degree $d$ (in this way a generic algorithm cannot classify elements on the basis of their degree), we find $c_{in}=\frac{d(1-2\rho)}{N(1-\rho)^2}$, 
$c_{out}=\frac{d}{N(1-\rho)}$.
The thresholds, in this case, are slightly different from the ones in Ref. \cite{BarbierIS} that were for graphs of fixed degree but the behavior is the same.
We restrict ourselves to the case of average degree $d=40$. In this case, for fixed degree, following Ref. \cite{BarbierIS}, 
the thresholds are $\rho_l(40)=0.138$, $\rho_{max}(40)=0.1273$, $\rho_s(40)=0.1231$.
We write the BP equations following the same reasoning of the ones for the clique problem, 
practically the equations are those in Eqs. (26),(27) of Ref. \cite{clusteringDecelle} for the general case of
clustering.
We numerically extract the threshold $\rho_l$ as the limit for the convergence of the BP equations to the planted solution once they are randomly initialized: $\rho_l(40)=0.135(1)$.
Analogously to the clique problem, we identify the static transition as the threshold at which the free energy of the planted solution, reached by BP initialized near enough to the
planted solution,
is equal to the free energy of the paramagnetic solution, reached by randomly initialized BP, finding $\rho_s(40)=0.1217(1)$, as shown in Fig \ref{Fig:f_IS}.
Please note that in the case of sparse IS the static threshold $\rho_s$ does not corresponds to the threshold $\rho_{max}$ for the maximal density of random IS, 
at variance with what happens in the clique problem, where we find that the static threshold $K_s$ numerically corresponds to the  
size $K_{max}$ of the largest naturally occurring clique.
We have that $\rho_s<\rho_{max}$. Even if for $\rho_s<\rho<\rho_{max}$ there exist random ISs, in this region the planted IS dominates the measure and can be reconstructed. 
In the dense limit $d\rightarrow\infty$, $\rho_{max}=\rho_s$ \cite{BarbierIS}, as it is expected because of the exact mapping into the dense clique problem.

\section{Parallel Tempering for the planted IS}
\begin{figure}
\centering
\includegraphics[width=0.48\columnwidth]{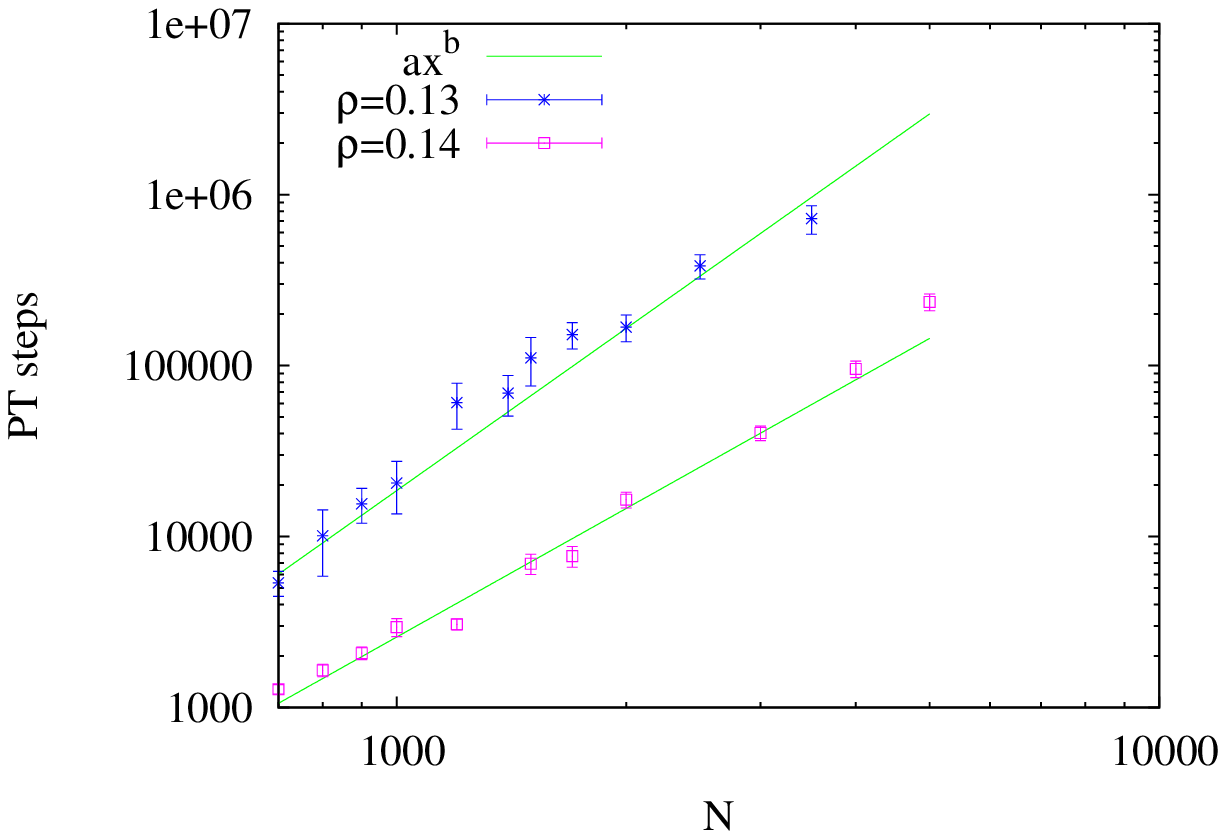}
\includegraphics[width=0.48\columnwidth]{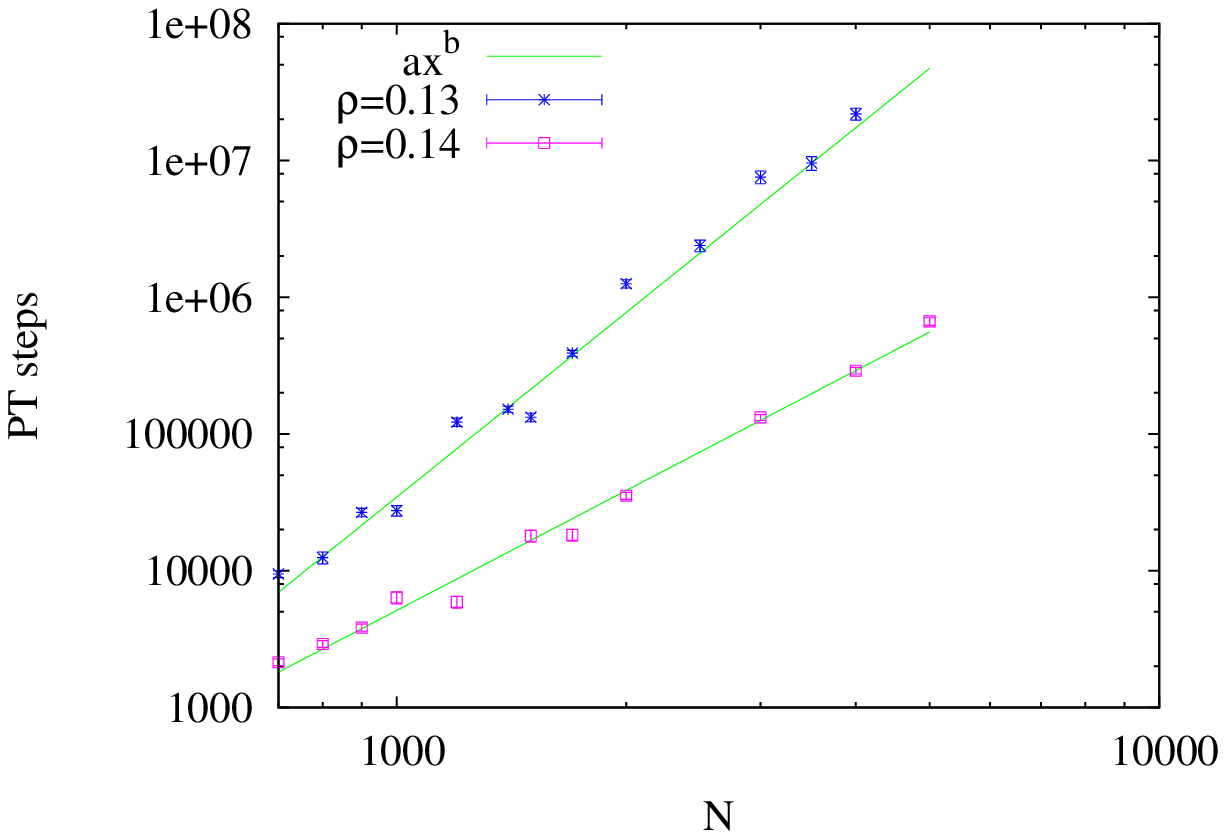}
\caption{\label{Fig:PT_IS}Average time of convergence (left) and 90th percentile time of convergence (right) for PT changing $N$
for $d=40$ and planted IS of density $\rho=0.13$ and $\rho=0.14$.
Times grow as a power-law with $N$. Each point is the average over $\sim 10^3-10^4$ realization of the graph.}
\end{figure}

Having identified the important transitions, we run PT in the hard region for the recovery of the planted solution.
As for the clique problem, we write the Hamiltonian associated to the posterior Bayes probability that is essentially the one in Eq. (8) of Ref. \cite{clusteringDecelle}.
A MC step takes time $O(dN)$. 
We introduce $n$ replicas at different inverse temperatures $\beta$: $\beta_i=1-i\cdot0.02$, $i\in[0,18]$.
We run PT at two different densities: $\rho=0.14$ that is in the easy phase, and $\rho=0.13$ that is in the hard phase for the reconstruction of the planted 
solution.
PT succeeds to find the planted solution, and the times are reported in Fig. \ref{Fig:PT_IS} as a function of $N$. 
For both $\rho=0.13$ and $\rho=0.14$ we have tried to fit data both with a polynomial $f(x)=ax^b$ and an exponential function $g(x)=c\exp(dx)$.
In both cases, the exponential fit has to be excluded, while times are well fitted with a polynomial growth.
The best-fit parameters for the exponent $b$ are:
$b(\rho=0.14)= 2.5(1)$,
$b(\rho=0.13)= 3.15(9)$.
Also the time of convergence of the $90\%$ of samples grows as a power-law with $N$ (see Fig. \ref{Fig:PT_IS}).

\section{Conclusions and perspectives}
Concluding, we have applied a standard method for the analysis of disordered models in statistical physics, the Parallel Tempering algorithm, to the planted clique problem.
The performances are quite surprising: It succeeds in finding the planted solution down to the information theoretical threshold in a time 
numerically compatible with a power-law in the size of the system. 
Moving to the planted IS sparse problem, that should be harder because thresholds become sharper,
the performances of PT still remains really good, succeeding in finding the correct solution in the hard region in polynomial time. 
The PT algorithm is a non-local algorithm because replicas of the system at different effective temperatures are flipped during the simulation.
When looking to the associated statistical mechanical problem, 
the hardness of the planted clique or planted IS problems relies on the presence of an extensive barrier in the free energy landscape between the correct planted solution and 
a wrong paramagnetic solution. However, when the temperature is added, this barrier can become smaller and eventually disappears. 
For this reason, in the PT algorithm, replicas at higher temperature can explore rapidly a larger space of configurations preventing from being trapped by the paramagnetic solution. 
This paper just shows numerical evidence that there could exist a polynomial algorithm for the clique problem and related problems. 
A crucial point in the PT algorithm is the choice of the number of replicas that should satisfy two important properties:
\begin{itemize}
 \item The $\beta$ associated to the last replica should be low enough to allow the system to explore the whole phase space, without trapping barriers. 
 \item The spacing in $\beta$ should be not so large: In this way, the difference in the energy associated to two near replicas could be small enough to permit the flipping with a non-null probability in Eq. (\ref{eq:flipping}).
\end{itemize}
For the analyzed sizes we have seen that $n=19$ is a good number to have both these properties. One could criticize that $n$ could grow with $N$: It could be possible
that a diverging number of replicas are needed in the thermodynamic limit to ensure the convergence for the PT algorithm to the planted solution. 
However for the analyzed sizes we verified that $n\simeq O(\log(N))$ is sufficient. Even if such a scaling of $n$ with $N$ is needed, the 
time of convergence of the algorithm will stay polynomial.

An analytical study of the 
performance of thermic algorithms is quite difficult. However, we think it could be of crucial importance for the determination of the real scaling of PT, 
and its application also for other optimization problems. 
The application of PT algorithm to the largest clique (and largest IS) problem on a Random Regular Graph is currently under study.

Let us finally point out the huge finite size effects of the planted clique problem for which the analytical thresholds for the static and the spinodal phase transitions
in the $N\to\infty$ limit, known in the literature, are so different from the real ones at finite but large sizes computed in this paper. This result suggests that 
the mathematical analysis of the $N\to\infty$ limit for such kind of problems should be complemented by a more deep study of the finite size behavior. A similar thing happens also 
for the Independent Set problem, for which there are a lot of mathematical results for the $d\to\infty$ limit, while at finite $d$ things are different, as 
found for example in Refs. \cite{BarbierIS}\cite{HartmannVC}. 

\section{Aknowledgment}
I thank Andrea Montanari for introducing me to the planted clique problem and for the suggestion to
look at the sparse Independent Set problem. I thank also Scott Kirkpatrick, Raffaele Marino, Federico Ricci-Tersenghi for very interesting discussions.


\begin{thebibliography}{100}
\bibitem{plantedclique} M. Jerrum, Random Struct. Algorithms 3 (1992) p.347. 
\bibitem{GM75}Geoffrey R Grimmett and Colin JH McDiarmid,On coloring random graphs, Mathematical Proceedings of the Cambridge Philosophical Society, vol. 77, Cambridge Univ. Press, 313–324 (1975).
\bibitem{Barak} Barak, Boaz, et al. "A nearly tight sum-of-squares lower bound for the planted clique problem." Foundations of Computer Science (FOCS), 2016 IEEE 57th Annual Symposium on. IEEE, 2016.
\bibitem{Montanari} Y. Deshpande, A. Montanari, "Finding hidden cliques of size $\sqrt{N/e}$ in nearly linear time." Foundations of Computational Mathematics 15.4 (2015): 1069-1128.
\bibitem{Montanari15} A. Montanari, J. Stat. Phys. \textbf{161}, 273 (2015).
\bibitem{hard1} Q. Berthet and P. Rigollet, arXiv preprint, arXiv:1304.0828 (2013).
\bibitem{hard2} Bruce  E.  Hajek,  Yihong  Wu,  and  Jiaming  Xu.
Computational  lower  bounds  for community  detection  on  random  graphs. In COLT,  899–928,  (2015).
\bibitem{hard3} Z. Ma and Y. Wu, Ann. Stat. 43 (2015) p.1089.
\bibitem{hard4} T.T. Cai, T. Liang and A. Rakhlin, Ann. Statist. \textbf{45}, 1403 (2017).
\bibitem{Karp} Karp, "Probabilistic Analysis of Some Combinatorial Search Problems.", in Algorithms and Complexity: New Directions and Recent Results, Academic Press, NY 1976.  
\bibitem{QuietPlanting} Florent Krzakala and Lenka Zdeborova, PRL \textbf{102}, 238701 (2009).
\bibitem{Castro_Ks} Arias-Castro, Ery, and Nicolas Verzelen. "Community detection in dense random networks." The Annals of Statistics 42.3 (2014): 940-969.
\bibitem{Matula1}D. Matula, On the complete subgraphs of a random graph, Combinatory Mathematics and its. Applications (Chapel Hill, 1970) 356–369
\bibitem{Matula2}D. W. Matula. The largest clique size in a random graph. Technical report, Department of Computer Science, Southern Methodist University, 1976
\bibitem{Bollobas}B. Bollobas and P. Erdos, Cliques in random graphs, Math. Proc. Camb. Phil. Soc. 80, 419 (1976).
\bibitem{clusteringDecelle} Decelle, A., Krzakala, F., Moore, C., Zdeborová, L., Asymptotic analysis of the stochastic block model
for modular networks and its algorithmic applications. Phys. Rev. E \textbf{84}(6), 066106 (2011).
\bibitem{Hu} D. Hu, P. Ronhovde and Z. Nussinov, Philosophical Magazine \textbf{92}, 406 (2012).
\bibitem{CS} F. Krzakala, M. Mézard, F. Sausset, Y. Sun, L. Zdeborová, Phys. Rev. X \textbf{2}, 021005 (2012).
\bibitem{PT} K. Hukushima and K. Nemoto, J. Phys. Soc. Jpn. \textbf{65}, 1604 (1996).
\bibitem{Reinforcement} A. Braunstein, R. Zecchina, Phys. Rev. Lett. \textbf{96}, 030201 (2006).
\bibitem{BarbierIS} J. Barbier, F. Krzakala, L. Zdeborova, and P. Zhang, The hard-core model on random graphs revisited, J. Phys.: Conf. Series 473, 012021 (2013)
\bibitem{HartmannVC} A. K. Hartmann and M. Weigt, J. Phys. A: Math. Gen. 36, 11069–11093 (2003).


\end{thebibliography}
\end{document}